# Compatibilidad del Argumento Cosmológico Kalam con la Ciencia Contemporánea

Desa, Carlos Bachiller en Física.

Universidad Nacional de San Antonio Abad Del Cusco, Av. la cultura 733, Cusco, Perú.

Apv. 28 de Julio E-7, Wanchaq, Cusco, Perú, 170735@unsaac.edu.pe, +51 940 824 978.

Compatibilidad del argumento Kalam con la ciencia actual.

Compatibilidad del argumento cosmológico Kalam con la ciencia actual.

**Resumen**

El argumento cosmológico Kalam, de carácter metafísico, sostiene que el universo tuvo una causa primera. Este trabajo analiza, mediante una revisión bibliografica de los avances de la física contemporánea, la compatibilidad de sus premisas.

La teoría cuántica muestra que los fenómenos cuánticos son indeterministas y no causales. Incluso en teorías deterministas, la causalidad carece de una definición clara. Además, los modelos cosmológicos, como la inflación cósmica, indican que los primeros instantes del universo no requieren una causa primera.

En conclusión, las premisas del argumento Kalam no son compatibles con la física contemporánea.

Compatibility of the Kalam Cosmological Argument with Current Science.

Abstract

The Kalam cosmological argument, of a metaphysical nature, asserts that the universe had a first cause. This study examines the compatibility of its premises through a bibliographic review of advancements in contemporary physics.

Quantum theory demonstrates that quantum phenomena are indeterministic and non-causal. Even in deterministic theories, causality lacks a clear definition. Moreover, cosmological models, such as cosmic inflation, suggest that the early stages of the universe do not require a first cause.

In conclusion, the premises of the Kalam argument are not compatible with contemporary physics.

## 1. Introducción

El argumento cosmológico Kalam, que es de carácter metafísico, se basa en dos premisas para afirmar que el universo tuvo un inicio. Actualmente, es defendido por teólogos como William Lane Craig (Craig, 2000). Sin embargo, estas premisas son supuestos sobre la realidad física, cuya compatibilidad con la física contemporánea será evaluada en este trabajo.

Este estudio se inspira en las objeciones presentadas por físicos y filósofos de la ciencia como Gustavo Romero (Romero, 2004) y Sean Carroll (Stewart, 2016). Se realiza una revisión literaria que considera avances clave en la física contemporánea. El teorema de Bell (Bell, 1964), junto con su más reciente verificación experimental (Herbst, T., y otros, 2015) basada en la interpretación de Copenhague (Faye & Jan, 2019), demuestra que la mecánica cuántica es inherentemente indeterminista y probabilística, lo que da lugar a fenómenos no causales. Además, en la relatividad general, que es una teoría determinista (Natsuume, 2015), la causalidad no está claramente definida debido a la naturaleza del espacio-tiempo.

Adicionalmente, las limitaciones del modelo del Big Bang han impulsado el desarrollo del modelo inflacionario (Baumann, 2009), que requiere describir estados previos al Big Bang. Por lo tanto, las premisas del argumento cosmológico Kalam no son compatibles con la física contemporánea.

Este trabajo se organiza de la siguiente manera: en la sección 2 se expone el argumento cosmológico Kalam. En la sección 3 se analiza la primera premisa, considerando el teorema de Bell, las implicaciones de las violaciones de las desigualdades de Bell y la causalidad en teorías deterministas. En la sección 4 se describe el modelo inflacionario y su necesidad de estados previos al Big Bang. Finalmente, la sección 5 presenta las conclusiones.

## 2. Argumento Cosmológico Kalam

El argumento cosmológico Kalam destaca por su simplicidad, ya que consta de dos premisas fáciles de entender y una conclusión. La formulación moderna más conocida de este argumento fue presentada por William Lane Craig en forma de un silogismo simple (Craig, 2000), el cual se expone de la siguiente manera:

1. Todo lo que comenzó a existir tiene una causa.
2. El universo comenzó a existir.

Por lo tanto,

El universo tiene una causa.

A continuación, se analiza la compatibilidad de estas dos premisas desde la física contemporánea.

3. Compatibilidad de la Primera Premisa del Argumento Cosmológico Kalam

La primera premisa del argumento cosmológico Kalam afirma que todo lo que existe tiene una causa, una afirmación basada en el principio de causalidad. Sin embargo, los fenómenos cuánticos han puesto en duda esta preposición. Durante el desarrollo de la teoría cuántica, surgieron dos interpretaciones principales. Por un lado, la interpretación de Copenhague (Faye & Jan, 2019) sugiere que los fenómenos cuánticos son intrínsecamente probabilísticos e indeterministas, lo que implica la existencia de fenómenos no causales en la naturaleza.

Para abordar estas cuestiones desde una perspectiva clásica, se consideró que la teoría cuántica era incompleta y que su aparente indeterminismo provenía de limitaciones epistemológicas. Esto llevó a la propuesta de las teorías de variables ocultas. Con base en ello, John S. Bell formuló su famoso teorema (Bell, 1964), según el cual "ninguna teoría física de variables ocultas locales puede reproducir todas las predicciones de la mecánica cuántica". Este teorema se fundamenta en las llamadas desigualdades de Bell, desarrolladas a partir de la estadística causal (Gill, 2022).

Diversos experimentos (Aceves Rodríguez, 2016), particularmente uno con gran rigor metodológico (Herbst, T., y otros, 2015), han demostrado violaciones de las desigualdades de Bell, confirmando la validez del teorema. Por lo tanto, se concluye que los fenómenos cuánticos son indeterministas, probabilísticos y que, efectivamente, existen fenómenos no causales en la naturaleza.

Por otro lado, la comprensión de la causalidad en las teorías clásicas modernas ha evolucionado significativamente con el desarrollo de la relatividad especial y general. En la relatividad general, es posible construir un espacio-tiempo con curvatura negativa, conocido como espacio-tiempo Anti-de Sitter (AdS). En este modelo, la coordenada temporal es periódica, lo que implica la existencia de un bucle temporal cuyo período está relacionado con el radio del espacio-tiempo (radio AdS). En este contexto, no hay una distinción clara entre una primera causa y un efecto final (Natsuume, 2015).

Sin embargo, las observaciones cosmológicas no han encontrado evidencia de que el universo posea una geometría negativa o un tiempo periódico, lo que hace problemático este modelo. Para resolver este problema, se extiende la coordenada temporal considerando una cobertura del espacio-tiempo, es decir, un radio AdS extremadamente grande. En este caso, el tiempo adquiere un comportamiento convencionalmente observado, solucionando así el problema.

Adicionalmente, el principio de relatividad establece que las leyes de la física son las mismas en todos los sistemas de referencia. Por ello, la causalidad solo tiene sentido dentro del cono de luz definido por el sistema inercial. Fuera de este cono, la noción de causalidad se vuelve ambigua, ya que, dependiendo del sistema de referencia, un evento puede considerarse causa o efecto de otro (Hacyan, 2004).

## 4. Compatibilidad de la Segunda Premisa del Argumento Cosmológico Kalam

La segunda premisa del argumento cosmológico Kalam sostiene que el universo tuvo un origen. Esta interpretación se basa en la visión clásica del "Big Bang", que tradicionalmente se entiende como el inicio del universo a partir de una gran explosión. Sin embargo, en los modelos contemporáneos, el "Big Bang" se refiere a una fase del proceso inflacionario en la que surgieron las primeras partículas fundamentales.

En la cosmología actual, el modelo teórico más aceptado es la teoría inflacionaria, que propone que el universo no se originó mediante una explosión violenta, sino a través de un proceso de expansión inflacionaria homogénea y acelerada (Tsujikawa, 2003). Esta teoría surgió principalmente para resolver problemas fundamentales de la cosmología clásica, como el problema del horizonte (Baumann, 2009). Según esta perspectiva, el universo atravesó una fase de expansión acelerada que dio lugar al "Big Bang", implicando la existencia de estados previos al evento tradicionalmente considerado como su origen.

Adicionalmente, se han desarrollado propuestas de teorías preinflacionarias que describen estados del universo anteriores al proceso inflacionario (Morais, J., Bouhmadi-López, M., Krämer, M., & Robles-Pérez, S., 2018). Estas teorías amplían la comprensión de la historia cósmica, sugiriendo que el universo no tuvo un origen, sino que existió en estados previos desconocidos antes del inicio de la inflación. Sin embargo, según nuestra comprensión actual del cosmos, solo se puede tener certeza acerca de los eventos que ocurrieron después del instante de Planck; antes de ese momento, todo sigue siendo especulativo.

## 5. Conclusiones

La naturaleza probabilística de la mecánica cuántica, con su indeterminación intrínseca, se manifiesta a través de fenómenos cuánticos no causales, como el decaimiento radiactivo. La validación del teorema de Bell respalda esta visión del mundo cuántico. En este contexto, la idea de causalidad parece ser un principio válido en el mundo clásico, pero no en el cuántico. Además, las teorías de la relatividad especial y general han demostrado que el concepto de causalidad no es absoluto, sino que depende del sistema de referencia. Asimismo, existen modelos cíclicos, compatibles con la relatividad general, que ofrecen explicaciones plausibles para el origen y la evolución del universo.

Los problemas de la cosmología contemporánea requieren modelos en los cuales el "Big Bang" clásico sea interpretado como una fase dentro de un proceso más complejo: el proceso inflacionario. Este proceso incluye fases preinflacionarias que permiten abordar cuestiones cosmológicas fundamentales. No obstante, los modelos previos al tiempo de Planck son meramente especulativos, y es necesario avanzar en estas investigaciones para lograr una comprensión más precisa de dichos procesos.

## Bibliografía